\documentclass[reqno,a4paper,final,12pt]{amsart}
\usepackage[british]{babel}
\usepackage{amssymb,amstext,amsthm,eucal,amsmath}
\usepackage[margin=1in]{geometry}
\usepackage{mathtools}
\usepackage{graphicx, caption}
\usepackage{wrapfig}
\usepackage{tikz}
\usetikzlibrary{arrows}
\usetikzlibrary{positioning}
\usetikzlibrary{calc}
\usetikzlibrary{shapes}
\usepackage{array}
\usepackage{amsfonts}
\usepackage{slashed}
\usepackage[utf8]{inputenc}
\usepackage[notref,notcite]{showkeys}
\usepackage[small]{eulervm}
\usepackage{tgpagella}
\usepackage{multicol}
\usepackage{float}
\usepackage{enumerate}
\usepackage{longtable,tabu}
\usepackage[section]{placeins}
\usepackage{hyperref}
\usepackage{ dsfont }
\usepackage{ upgreek }
\usepackage{ mathrsfs }
\usepackage{braket}
\usepackage{gensymb}
\usepackage{subcaption}
\usepackage{listings}

\tikzset{every picture/.style={line width=1pt}} 

\definecolor{lightgray}{gray}{0.9}
\newcommand{\inlinecode}[2]{\colorbox{lightgray}{\lstinline[language=#1]$#2$}}

\hypersetup{
  pdftitle   = {DFORMPY: A Visualising and Calculus Differential Forms Interactive Python Library},
  pdfkeywords = {differential forms, vector fields, exterior derivative, interior derivative, Hodge, Wedge, curl, divergence},
  pdfauthor  = {Moustafa Gharamti},
  pdfcreator = {\LaTeX\ with package \flqq hyperref\frqq}
}

\ifx\Sp\UnDeFiNeD
\newcommand{\Sp}{\mathrm{Sp}}
\else
\renewcommand{\Sp}{\mathrm{Sp}}
\fi

\definecolor{orange}{rgb}{0.9,0.45,0}
\definecolor{green}{rgb}{0,0.5,0}
\newcommand{\MUNCH}[1]{\relax}

\allowdisplaybreaks[1]
\begin{document}
\title{DFORMPY: A Python Library for visualising and zooming on differential forms}
\author[G.J.K.]{Moustafa Gharamti$\dagger$, Maciej Jarema$\circ$, Samuel Kirwin-Jones$\star$ \\
School of Physics and Astronomy \\
University of Nottingham
\\ $\dagger$ \href{mailto:moustafa.gharamti@nottingham.ac.uk}{\fontsize{10}{12}\selectfont moustafa.gharamti@nottingham.ac.uk}
\\ $\circ$ \href{mailto:macusjarema@gmail.com}{\fontsize{10}{12}\selectfont macusjarema@gmail.com}
\\ $\star$ \href{mailto:1999samkj@gmail.com}{\fontsize{10}{12}\selectfont 1999samkj@gmail.com}}
\begin{abstract}
We present the v1.0.1 release of DFormPy, the first Python library providing an interactive visualisation of differential forms. DFormPy is also capable of exterior algebra and vector calculus, building on the capabilities of NumPy and matplotlib. This short paper will demonstrate the functionalities of the library, briefly outlining the mathematics involved with our objects and the methods available to the user. DFormPy is an open source library with interactive GUI released under MIT license at \url{https://github.com/MostaphaG/Summer_project-df}.
\end{abstract}
\maketitle
\vspace{-1cm}
\tableofcontents

\section{Introduction}
Differential forms are objects allowing the study of differential geometry, a branch of mathematics that underpins much of modern physics. They first make their appearance in the latter stages of undergraduate physics courses in topics such as relativity and electromagnetism. Many undergraduate physicists will never witness the power and value of differential forms and exterior algebra. This is a great shame, given they generalise the notion of a vector field to any dimensionality, utilising cleaner notation and a more intuitive geometrical meaning.

To accompany the lack of teaching of differential forms in physics departments, there is little available to the community of scientific programmers, which hinders the prospect of self education in this field. As a result, we have attempted to create the tools required to learn the basics of differential forms, culminating in the first edition Python library called DFormPy.

We are aware of only one other work that attempted at computationally visualising differential forms, namely ``Vector Field Analyser'' (VFA) \cite{holland2000interactive, Kawski2004}. VFA is a JAVA code and used to run via JAVA applet on web browsers \cite{VFA2020}. It focuses on visualising vector fields and their calculus and plots covariant vectors (`stacks'). Our work, in addition to making their achievements more accessible (through Python), far expands on the idea of covariant vectors ($1$-forms) and their governing differential geometry.

Our library provides the capability to visualise and manipulate differential forms and vector fields in Python. This version allows full plotting and exterior algebra operations in two-dimensions (which we will be extending to three-dimensions in the near future) and we aim to make the functions as user friendly as possible to both Python beginners and regular users of matplotlib and NumPy libraries.

The library is also capable of demonstrating the local variations of the curl and divergence of vector fields, as well as their total derivative. Also, it helps the user understand the geometric meaning of the exterior and interior derivatives, the Hodge star operation, the wedge product and the involvement of the metric of curved two dimensional manifolds.

All of the above have far-reaching applications in demonstrating, teaching and testing concepts commonly used in General Relativity, Electromagnetism and Linear Algebra. Combined with our GUI, the users are also equipped with the tools to study and test the behaviour of differential equations, line integrals and area integrals, thus providing a platform to interact with concepts in vector calculus.



In this paper, firstly we give a concise review of the differential forms construction and their algebra. Next, `DFormPy objects' are introduced and their respective methods are outlined using some example plots. We then elaborate on some of the technical details of these methods before finally utilising them in physically useful examples.

\section{Mathematical preliminaries}
In this section, we briefly setup the mathematical definitions of differential forms and their exterior algebra. For detailed discussion and practices we recommend that the reader checks the literature such as \cite{fortney2018visual, do1998differential, frankel2011geometry}. 

\subsection{Differential $p$-forms} 
Let $V$ be a vector space over $\mathbb{R}$ with dimension $n$. The dual space $V^*$ is a vector space of forms with dimension $n$, such that if ${\pmb e}_i$ and ${\pmb {\theta}}^i$ are the bases of the $V$ and $V^*$ respectively, then 
\begin{equation}
{\pmb {\theta}}^i(\pmb e_j) \coloneqq \delta^i\,_j
\end{equation}
The vector space of differential $p$-forms is denoted by $\wedge^pV$, where $p$ is a positive integer.  $\wedge^pV$ is the $p^{th}$ completely antisymmetric tensor power of $V^*$.  In particular, $\wedge^1V=V^*$, $\wedge^pV=0$ for $p>n$ and by convention $\wedge^0V= \mathbb R$.\\

\noindent If ${\pmb\alpha} \in \wedge^pV$, then 
\begin{equation}
\pmb\alpha(\dots, \pmb e_i, \dots, \pmb e_j, \dots)=-\pmb\alpha(\dots, \pmb e_j, \dots, \pmb e_i, \dots)
\end{equation}
where $p$ is referred to as the degree of $\pmb\alpha$. A typical representation of $\pmb\alpha$ is given by
\begin{equation}
\pmb\alpha=\frac{1}{p!}\sum_{i_1,\dots,i_p}\alpha_{i_1\dots i_p}{\pmb \theta}^{i_1}\wedge\dots\wedge{\pmb \theta}^{i_p}.
\end{equation}
For instance, if ${\pmb\alpha} \in \wedge^1\mathbb R^2$ and ${\pmb\omega} \in \wedge^2\mathbb R^2$, then
\begin{align}
\pmb\alpha &=\alpha_1(x,y) dx+\alpha_2(x,y) dy,\\
\pmb\omega &=\frac{1}{2}[\omega_{12}(x,y) dx\wedge dy+\omega_{21}(x,y) dy\wedge dx]
\end{align}
where $dx$ and $dy$ denote the basis of $V^*$.

\subsection{$V^*$ as tangent space}
The abstract dual vector space where we define differential forms can be the dual tangent space at a point on a manifold. For DFormPy the manifold is ${\mathbb R}^2$, so our differential $p$-forms are in $\wedge^pT^*{\mathbb R}^2$. Since the tangent space to a vector space, seen as a manifold, at any point is isomorphic to the vector space itself \cite{bellaiche1996tangent}, we have
\begin{equation}
T^*{\mathbb R}^2 \cong {\mathbb R}^2,
\end{equation} 
and so we will always refer to $\wedge^pT^*{\mathbb R}^2$ as $\wedge^p{\mathbb R}^2$.

\subsection{$p$-Forms exterior algebra} 
In the following, we give the mathematical definitions and main properties of the algebra of differential forms.

\subsubsection{Wedge product $\wedge$} \label{wedgef}In this operation we take the exterior product between differential forms. The result is a differential form of degree equal to the sum of the degrees of all the differential forms involved.\\
If $\pmb \alpha \in \wedge^pV$ and $\pmb \omega \in \wedge^qV$, then
\begin{eqnarray}
\wedge: \wedge^pV\times\wedge^qV&\to&\wedge^{p+q}V\nonumber\\
(\pmb\alpha,\pmb\omega)&\to&\pmb\alpha\wedge\pmb\omega
\end{eqnarray}
$\wedge$ is bilinear, associative and graded commutative. The last property meaning:
\begin{equation}
\pmb\alpha\wedge\pmb\omega = (-1)^{pq}\pmb\omega\wedge\pmb\alpha
\end{equation}

\subsubsection{Exterior derivative $d$}\label{extd} This operation maps $p$-forms onto $p+1-$forms by performing derivatives of the forms coefficients w.r.t.~their variables and wedging the result linearly by the corresponding bases:
\begin{eqnarray}
d: \wedge^pV&\to&\wedge^{p+1}V\nonumber\\
\pmb\alpha&\to&d\pmb\alpha
\end{eqnarray}
where 
\begin{equation}
d\pmb\alpha =\frac{1}{p!}\sum_{i_1,\dots,i_p}\sum_{j=1}^{n}\frac{\partial\alpha_{i_1\dots i_p}}{\partial x^j}dx^j\wedge{dx}^{i_1}\wedge\dots\wedge dx^{i_p}.
\end{equation}
In particular, if $\phi \in \wedge^0{\mathbb R}^2$ and ${\pmb \alpha} \in \wedge^1{\mathbb R}^2$, then $d\phi  \in \wedge^1{\mathbb R}^2$ and $d\pmb\alpha  \in \wedge^2{\mathbb R}^2$ and are given by
\begin{equation}
d\phi = \frac{\partial \phi}{\partial x}dx +  \frac{\partial \phi}{\partial y}dy \quad \text{and} \quad d{\pmb \alpha} = (\frac{\partial \alpha_2}{\partial x} - \frac{\partial \alpha_1}{\partial y})dx\wedge dy 
\end{equation}
$d$ is linear and satisfies graded Leibniz rule, i.e. 
\begin{equation}
d(\pmb\alpha\wedge\pmb\omega) = (d\pmb\alpha)\wedge\pmb\omega+(-1)^p\pmb\alpha\wedge d\pmb\omega, \quad \pmb\alpha\in\wedge^pV.
\end{equation}
\subsubsection{Interior derivative $\iota_{\pmb v}$} \label{intd} In this operation we act on differential form w.r.t.~to a vector field. This operation lowers the degree of the differential form by one unit.\\ 
Let $\vec v~=~\sum_jv^j{\pmb e}_j \in V$, then 
\begin{eqnarray}
\iota_{\vec v}: \wedge^pV&\to&\wedge^{p-1}V\nonumber\\
\pmb\alpha&\to&\iota_{\vec v}\pmb\alpha
\end{eqnarray}
where
 \begin{align}
\iota_{\vec v}\pmb\alpha =\frac{1}{p!}\sum_{i_1,\dots,i_{p-1}}&\Big[\sum_{j}v^j\alpha_{ji_1\dots i_{p-1}}\Big]{\pmb\theta}^{i_1}\wedge\dots\wedge {\pmb\theta}^{i_{p-1}}.
\end{align}
In particular, if $\pmb\alpha \in \wedge^1V$ then $\iota_{\vec v}\pmb\alpha\in\mathbb{R}$.\\
$\iota_{\vec v}$ is linear in ${\vec v}$ i.e.
\begin{align}
\iota_{{\vec v}+{\vec w}} &= \iota_{\vec v}+\iota_{\vec w}, \quad {\vec v}, {\vec w} \in V \\
\iota_{a\vec v}&=a\iota_{\vec v}, \quad a \in \mathbb R. 
\end{align}
$\iota_{\vec v}$ satisfies graded Leibniz rule, i.e. 
\begin{equation}
\iota_{\vec v}(\pmb\alpha\wedge\pmb\omega) = (\iota_{\vec v}\pmb\alpha)\wedge\pmb\omega+(-1)^p\pmb\alpha\wedge\iota_{\vec v}\pmb\omega, \quad \pmb\alpha\in\wedge^pV.
\end{equation}
Moreover, one can show that 
\begin{equation}
\iota_{\vec v}\iota_{\vec w}+\iota_{\vec w}\iota_{\vec v}=0.
\end{equation}
\subsubsection{ The Hodge $\star$} \label{hdgfrm}This operation acts on differential forms as a linear map 
\begin{eqnarray}
\star: \wedge^pV&\to&\wedge^{n-p}V\nonumber\\
\pmb\alpha&\to&\star\pmb\alpha
\end{eqnarray}
where
 \begin{align}
\star\pmb\alpha =\frac{1}{(n-p)!}\sum_{i_{p+1},\dots,i_n}&\Big[\frac{1}{p!}\sum_{i_1,\dots,i_p}\varepsilon_{i_1\dots i_n}|\det g_{kl}|^{1/2}\nonumber\\
&\times\sum_{j_1,\dots,j_p}\pmb \alpha_{j_1\dots j_p}g^{i_1j_1}\dots g^{i_pj_p}\Big]{\pmb\theta}^{i_{p+1}}\wedge\dots\wedge {\pmb\theta}^{i_n}.
\end{align}
$g_{kl}$ is the metric of the manifold over which the vector space $V$ is defined, and equal to the inner product of the $V$ bases, $g_{kl} = \braket{\pmb e_k,\pmb e_l}$. $\varepsilon_{i_1\dots i_n}$ is the $n-$dimensional Levi-Civita symbol. \\
For instance, if ${\pmb\alpha} \in \wedge^1\mathbb R^2$ and  ${\pmb\omega} \in \wedge^2\mathbb R^2$ such that ${\pmb\alpha} = \alpha dx $ and ${\pmb\omega} =\omega dx\wedge dy$, then
\begin{align}
\star \pmb\alpha=\alpha dy \in \wedge^1{\mathbb{R}}^2 \quad \text{and}\quad \star \pmb\omega&=\omega \in \mathbb R \nonumber
\end{align}

\section{DFormPy objects}
As discussed in the previous section and in more details in \cite{fortney2018visual} \cite{do1998differential} and \cite{frankel2011geometry}, vector fields and differential forms are different geometrical objects and thus will behave in unique ways under the operations described previously. 

%

DFormPy provides classes which are used to create instances. This way, it is simple and intuitive for the user to manipulate the forms, while ensuring that the operations made available to the user are suitable for any particular instance. Available classes in $\mathbb{R}^{2}$ are vector field, $0$-form, $1$-form and $2$-form.

Plotting, customisation and mathematical operations are performed using methods which belong to the class. Some of these included methods are capable of returning a different type instance after being called (e.g. a \inlinecode{Python}{.hodge( )} method acted on the $2$-form, will return a $0$-form). Through this, the differences and connections (via exterior algebra) between objects are made clear, making Python an excellent resource to explore differential forms.\\

%
\noindent An overview of these classes and connections between them is shown below:
\small
\begin{center}
\hspace*{-1.5cm}%
\begin{tikzpicture}[
  node distance = 2cm,
info/.style = {rectangle, ultra thick, draw=green!50, minimum width=5cm,
             minimum height=3cm,  label=above: {\bf DFormPy}},
obj/.style = {rectangle, ultra thick, draw=green!50, fill=blue!20, minimum width=1.2cm,
             minimum height=0.8cm, label=right: ~ ~Objects},             
rect/.style = {rectangle, ultra thick, draw=green!50, fill=blue!20, minimum width=2.5cm,
             minimum height=2cm, text centered},
  line/.style = {-latex'},
  meth/.style = {-stealth, thick, draw=black, label=right: ~ ~Methods},
  ]
 
\node[rect] (R1) {Vector Field};
\node[info, above =  0.5cm of R1, xshift=-5cm, yshift = -2cm  ] (R5) {};
\node[obj, above =  0.2cm of R1, xshift=-6cm] (R6) {};
\node[rect, below =  2cm of R1, xshift=-6cm  ] (R3) {$0$-form};
\node[rect, below = 2cm of R1] (R2) {$1$-form};
\node[rect, below = 2cm of R1, xshift=+6cm  ] (R4) {$2$-form};
  
  
\draw [-to](-6.5,0.4) -- (-5.3,0.4) node[pos=0.5, right] {~~~~~~ Methods};
\draw [line] (R1) -- ($(R1.center) + (0.7,-1)$) -- ($(R2.center) + (0.7,+1)$) node[pos=0.5, right] {.covariant( )} (R1);  
\draw [line] (R2) -- ($(R2.center) + (-0.7,+1)$) -- ($(R1.center) + (-0.7,-1)$ ) node[pos=0.5, left] {.contravariant( )} (R2);  
\draw [line] (R2) -- ($(R2.east) + (0,-0.6)$) -- ($(R4.west) + (0,-0.6)$) node[pos=0.5, below] {.ext\_d( ), .wedge( )} (R2);  
\draw [line] (R4) -- ($(R4.west) + (0,+0.6)$) -- ($(R2.east) + (0,+0.6)$) node[pos=0.5, above] {.interior\_d( )} (R4);  

\draw [line] (R3) -- ($(R3.east) + (0,-0.6)$) -- ($(R2.west) + (0,-0.6)$) node[pos=0.5, below] {.ext\_d( ), .wedge( )} (R3);  
\draw [line] (R2) -- ($(R2.west) + (0,+0.6)$) -- ($(R3.east) + (0,+0.6)$) node[pos=0.5, above] {.interior\_d( )} (R2);  
\draw [line] (R4) -- ($(R4.center) + (0,-4)$) -| ($(R3.center) + (0,-1)$ ) node[pos=0.25, below] {.hodge( )} (R3);  
\draw [line] (R3.300) |- ($(R3.300)  + (0,-1.9)$) -| ($(R4.240)$ )node[pos=0.25, below] {.hodge( ), .wedge( )}(R4) ;

\draw [->] (R2.250) arc (135:405:5mm) node[pos=0.5, below] {.zoom( ), .hodge( ), .wedge( )} (R2);
\draw [->] (R1.340) arc (225:495:5mm) node[pos=0.5, right] {.zoom( ), .deriv( ), .div(), .curl()} (R1); 
\draw [->] (R4.340) arc (225:495:5mm) node[pos=0.7, above] {~~~~.zoom( )}node[pos=-1, below] {~~~~~.wedge( )} (R4); 
\draw [->] (R3.160) arc (45:315:5mm) node[pos= 0.9, below] {.wedge( )~~~~~~~~~~~~~~~~~} (R3); 
 
\end{tikzpicture}\\~\\

\end{center}
\normalsize

In the following~\ref{sub1F}~to~\ref{sub0F}, we outline the objects and then in ~\ref{methi}~to~\ref{cotocontra}, describe the methods that can be used for each, with examples. Throughout, we follow a naming convention of $u$ and $v$ referring to the $x$ (or $dx$) and $y$ (or $dy$) components, respectively. The 2-form and 0-form scaling functions are consistently referred to as $w(x,y)$ and $\phi(x,y)$, respectively.


\subsection{$1$-Form}
\label{sub1F}
$1$-forms are covariant (dual) vectors. They effectively contract vectors (arrows), returning a number. One way to envisage this process is to think of the $1$-form as a set of sheets that lie perpendicular to a common axis and have a given orientation \cite{weinreich2021geometrical}. A vector that exists in the same position as this `stack' will pierce a certain number of these sheets, depending on its magnitude and relative orientation to the axis. The number of sheets pierced is the inner product of the $1$-form and the vector.
DFormPy $1$-forms adopt this representation and plot stack fields with a relative density and orientation matching the magnitude and orientation of an equivalent vector field. For that, the coordinates for the end-points for the stack sheets are calculated from the magnitude and direction of the supplied $dx$ and $dy$ components at each grid position.

\subsubsection{Methods} The $1$-form instance can be created by calling \inlinecode{Python}{dformpy.form_1( )}, with $x$, $y$, $u$ and $v$ grids (\inlinecode{Python}{numpy.ndarrays}) provided as input parameters. Users may additionally provide the strings for their $1$-form components, or provide them using the method \inlinecode{Python}{.give_eqn( )} once the instance has been created. All visual properties can be changed by acting on the $1$-form instance with one of our customisation methods listen in Appendix \ref{A1}.
It can be plotted, zoomed onto at a particular position and changed into a vector field instance via the inverse of the metric. It can also be acted on with a hodge, wedge (with a $0$-form and a $1$-form), interior derivative (returning a $0$-form instance), and the exterior derivative (returning a $2$-form instance). All of the above operations can be completed analytically and/or numerically, depending on user's needs or preference.

\subsubsection{Example}
In the following example, we show how one can import DFormPy and we present an example using the $1$-form instance.
\begin{figure*}[ht]
    \centering
    \begin{subfigure}[t]{0.3\textwidth}
        \centering
        \fbox{\includegraphics[height=4. in, width=3.2 in]{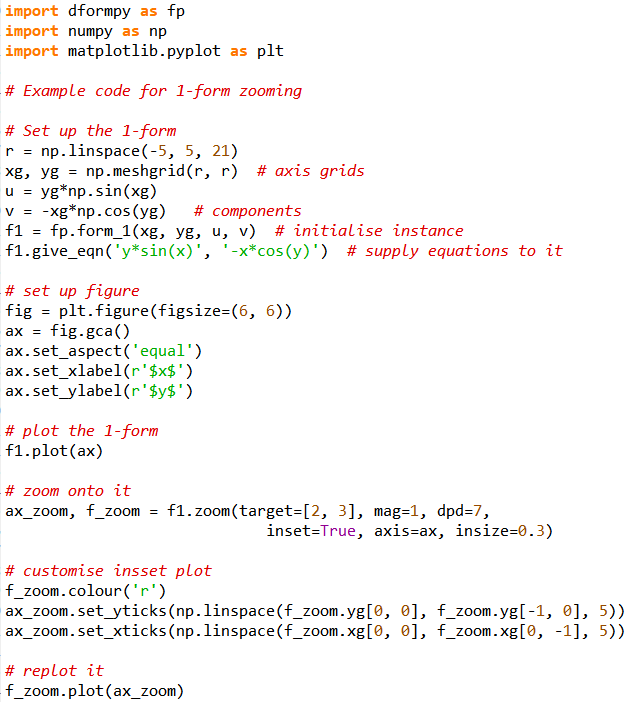}}
        \caption{Example code 1}
        \label{form_1_zoom_code}
    \end{subfigure}%
    \begin{subfigure}[t]{1\textwidth}
        \centering
        \includegraphics[height=4.1 in]{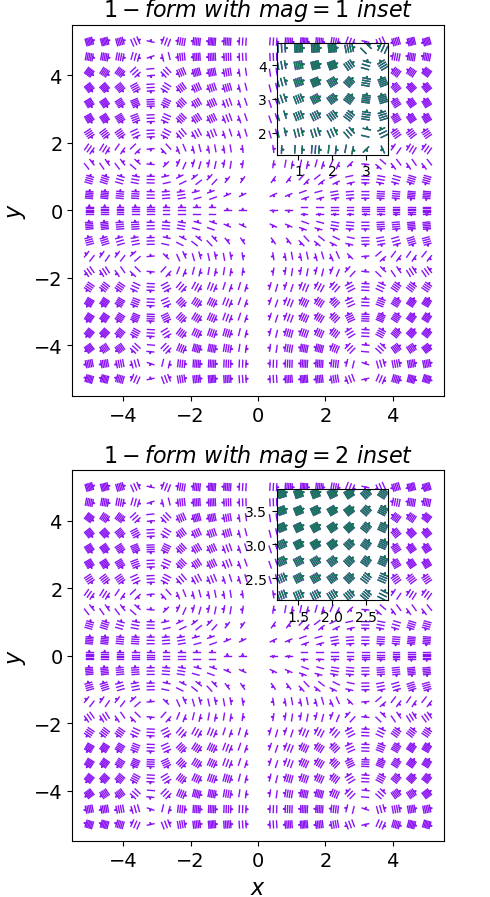}
        \caption{$1$-form and insets}
        \label{fig:form1_and_zooms}
    \end{subfigure}
    \caption{\fontsize{8}{10}\selectfont Plotting and zooming $1$-forms in $\mathbb{R}^2$. The purple plots \ref{fig:form1_and_zooms} were created using Cartesian grids with dimension $31 \times 31$ and a maximum of 6 sheets per stack with $u(x, y) = ysin(x)$, $v(x, y)=-xcos(y)$. Zoomed $1$-forms have grids with dimension $7 \times 7$ and the insets are centred on $(2,3)$ with \inlinecode{Python}{insize} of $0.3$. Magnification increases from 1 to 2 going from upper to lower plot; the zoomed form components become approximately constant.}
\end{figure*}
%


\subsection{Vector field (VF)}
These are elements of the vector space and usually referred to as contravariant vectors. On $\mathbb{R}^2$, a vector field is typically represented as 
\begin{equation}
\vec{F}(x,y) = u(x,y)\hat{x} + v(x,y)\hat{y},
\end{equation}
where $u(x,y)$ and $v(x,y)$ are scalar functions on ${\mathbb R}^2$.
DFormPy uses quiver plots from matplotlib and offers the ability to carry out differential operations, whilst allowing for easy customisation and modification.

\subsubsection{Methods}
A VF instance can be established using \inlinecode{Python}{dformpy.vector_field( )}. It can be plotted using the method \inlinecode{Python}{.plot( )}, which utilises \inlinecode{Python}{.quiver()} from matplotlib. The class instance stores the customisation information, which can be modified via methods listed in Appendix \ref{A2}. The vector field can be zoomed on using \inlinecode{Python}{.zoom( )}. Similarly, the method \inlinecode{Python}{.deriv( )} allows the user to create a new, magnified field representing the total derivative, which shows how the field changes in the local region about a specified target position. The derivative field encodes both how the VF is rotating and expanding at this point. These components are extracted using the methods \inlinecode{Python}{.curl( )} and \inlinecode{Python}{.div( )} respectively. 


The vector field (contravariant) can also be changed into a $1$-form (covariant) instance (described in \ref{cotocontra}), via the metric, using a method called \inlinecode{Python}{.covariant( )}.

\subsubsection{Example}
Here, we show an example plot of a vector field instance, and inspect the result by zooming and taking the derivative.
\begin{figure}[ht]
   	\centering
    \includegraphics[height=3.2 in]{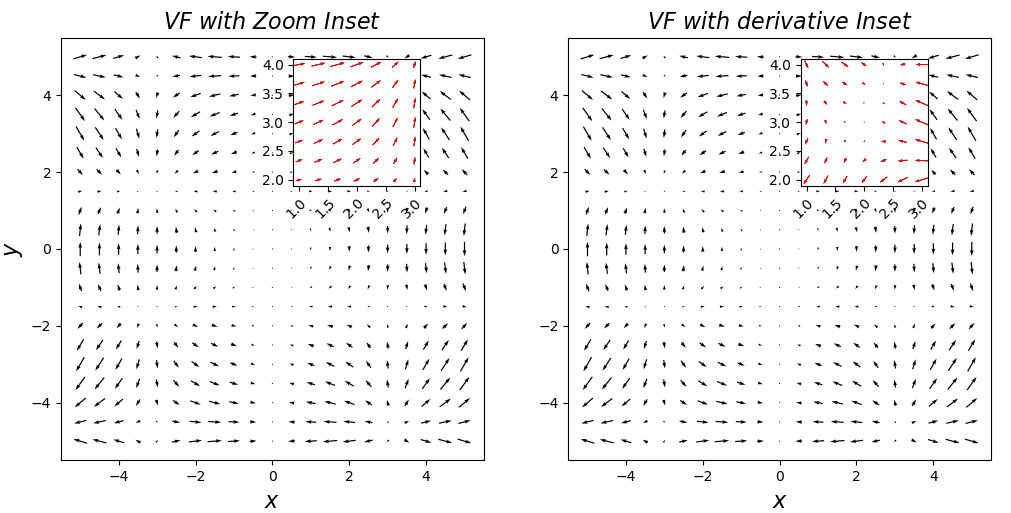}
    \label{VF_zoom}
    \caption{\fontsize{8}{10}\selectfont Zoom and derivative on vector field object using inset plots. The same vector field (with $u(x, y) = ysin(x)$, $v(x, y)=-xcos(y)$) has been plotted on both axes. Both figures create insets with magnification of $1.5$ at target position $(2, 3)$. Left figure shows an inset with the zoomed VF, using \inlinecode{Python}{VF.zoom( )}, whereas the right figure includes an inset with the total derivative (or Lie derivative), using \inlinecode{Python}{VF.deriv( )}}. 
\end{figure}

\subsection{$2$-Form}
\label{sub2form}
$2$-Forms are represented by superpositions of the sheets of $1$-forms. In ${\mathbb R}^2$, a general $2$-form can be expressed as $\Omega = w(x,y) dx \wedge dy$ where $w(x,y)$ is a real function on ${\mathbb R}^2$.
Since this version of DFormPy works in 2D, we can represent these objects by plotting a grid of superposed stacks, and representing the sign at a given grid point using colour. $2$-forms are to be input in terms of the scaling function $w(x,y)$ for $dx\wedge dy$ (counter-clockwise being positive). To define a $2$-form in $dy\wedge dx$, a minus sign must be added to $w(x,y)$.

Default colouring is red for counter-clockwise (positive $dx\wedge dy$, out of page), and blue for clockwise (negative $dx\wedge dy$, into page) and grey for zero magnitude. Variation is shown by the change in the number of squares from one position to the next. The stack density, just as for $1$-forms, is calculated relatively at each grid point (i.e. relative to the variation in the magnitude of $w(x,y)$ across the plotting region.) 

\subsubsection{Methods}
A $2$-form instance is created using \inlinecode{Python}{dformpy.form_2( )}. Just as for the other objects, the user provides the $x$ and $y$ grid values and a grid representing the values of $w(x,y)$, using NumPy (\inlinecode{Python}{numpy.ndarray}). The user may additionally provide the string equation for $w(x,y)$ if they wish to use analytical methods.
The $2$-form can be plotted using \inlinecode{Python}{.plot()}, and customisations to its visualisation can be applied with methods listed in Appendix \ref{A4}.
\inlinecode{Python}{.zoom( )} allows the user to create a new $2$-form instance and its inset plot, localised to a target location with a specified magnification and grid point density.
\inlinecode{Python}{.hodge( )} takes the Hodge star of the current $2$-form object, returning the $0$-form given by the function $w(x,y)$. 
\inlinecode{Python}{.interior_d( )} takes the interior derivative of the $2$-form with respect to a user specified vector field.

\subsubsection{Example}
Here, we present an example use of our $2$-form instance.
\begin{figure}[H]
		\centering
	\includegraphics[scale=0.5]{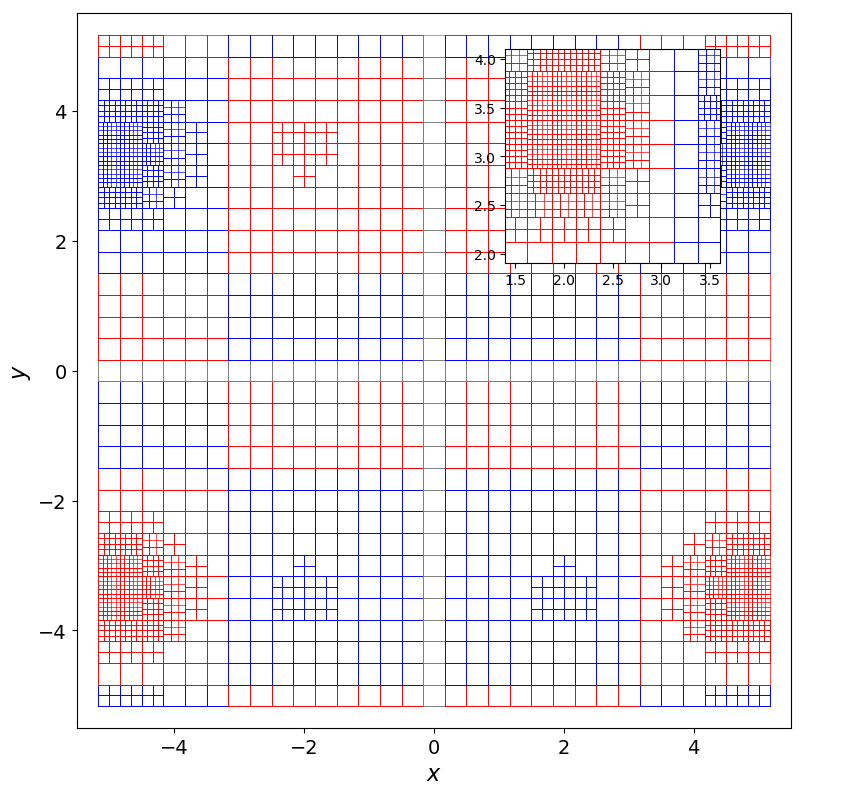}
	\caption{\fontsize{8}{10}\selectfont 
	Plotting and zooming $2$-forms in $\mathbb{R}^2$. Shows the variation in the $2$-form  defined by $w(x, y)=-xysin(x)cos(y)$. Zoom inset has target location of ($2.5, 3$) and magnification of $1.5$.
	}
	\label{fig:form2_and_zooms}
\end{figure}


\subsection{$0$-Form}
\label{sub0F} 
In ${\mathbb R}^2$, $0$-forms are duals of scalar fields and belong to $\wedge^{0}\mathbb{R}^2$. These are functions $\phi(x,y)$ that assign a single number to every position $(x, y)$ in $\mathbb{R}^2$. The $0$-form DFormPy instance is plotted as a contour map by \inlinecode{Python}{.contour()} from matplotlib, where each contour curve indicates a level line.

\subsubsection{Methods} The $0$-form instance can be created by calling \inlinecode{Python}{dformpy.form_0()}, providing $x$ and $y$ grids and a grid for the scalar function $\phi(x,y)$. In the same way for the other DFormPy objects, equations are provided as instance parameters or using \inlinecode{Python}{.give_eqn()}. Plot customisations, are stored within the instance, can are modified by methods listed in Appendix  \ref{A3}.
Users can also compute the exterior derivative, \inlinecode{Python}{.ext_d()} or \inlinecode{Python}{.num_ext_d()} for numerical calculation, or act on it with Hodge operator, \inlinecode{Python}{.hodge( )}, which returns a $2$-form instance.

\section{DFormPy methods}
Here, we describe how to use the methods available for all objects and their implementation in code. We also present example plots demonstrating exterior algebra in DFormPy.

\subsection{Plotting}\label{methi}
Once an instance of any of our classes has been created, it can be plotted using a method called \inlinecode{Python}{.plot(axis)}, where axis is the name assigned to the desired matplotlib axis. For example, one could create an axis using:\\
\hspace*{1cm}\inlinecode{Python}
{import matplotlib.pyplot as plt}\\
\hspace*{1cm}\inlinecode{Python}
{       figure = plt.figure()}\\
\hspace*{1cm}\inlinecode{Python}
{	axis = figure.gca()}\\
%
As previously mentioned, customisations to plots are stored in the instances themselves and can thus only be changed before \inlinecode{Python}{.plot( )} is called. The full list and descriptions of these can be found in Appendix \ref{appendix_custom}.

\subsection{Zooming}
To zoom in on a DFormPy object (excluding the $0$-form), call the \inlinecode{Python}{.zoom( )} method. This creates a new instance of the same type as the parent field. The parameters are:
\begin{itemize}
\item target: origin of new field, tuple of $x$ and $y$ coordinates (default = (0,0))
\item mag: level of zooming, positive float $\geqslant 1$ (default = 2) 
\item dpd: dimension of new $n \times n$ field, positive integer (default = 9)
\item inset: whether or not to create inset axis for plotting the zoom field, boolean (default is True) 
\item axis (for inset=True only): parent axis to plot the inset on
\item insize (for inset=True only): size of the inset axis as a fraction of the parent axis, positive float $\leqslant 1$ (default is $0.3$).
\end{itemize}

The method will return the new DFormPy object, and the matplotlib inset axis (when inset is True), to allow for customisation. The same format is used for \inlinecode{Python}{.deriv( )}, \inlinecode{Python}{.div( )}, and \inlinecode{Python}{.curl( )} methods, when working with vector fields.
Creation of zoomed fields require evaluation of the field components at the new target location. Consequently, in order to use the method, the user must provide the equations of their parent field components, either when they first create the parent instance, or using the \inlinecode{Python}{.give_eqn( )} method.

\subsection{Exterior derivative}
As discussed in \ref{extd}, the exterior derivative method can be used on $0$-forms to create $1$-forms, on $1$-forms to create $2$-forms, and returns zero when used on $2$-forms. To compute analytically, use the method \inlinecode{Python}{.ext_d( )}, ensuring the field components have been provided. To compute numerically, use \inlinecode{Python}{.num_ext_d( )}. For most fields, these two methods give very similar $2$-forms, but in general the analytical method gives a more accurate result. No additional parameters are required for these methods. The calculation of the analytical exterior derivative is done using SymPy expressions. Numerically, we use \inlinecode{Python}{numpy.gradient( )}, which uses a finite difference method to estimate the derivatives in vertical and horizontal directions.
\begin{figure*}[ht]
    \centering
    \begin{subfigure}[t]{0.3\textwidth}
        \centering
        \fbox{\includegraphics[height=4.35 in]{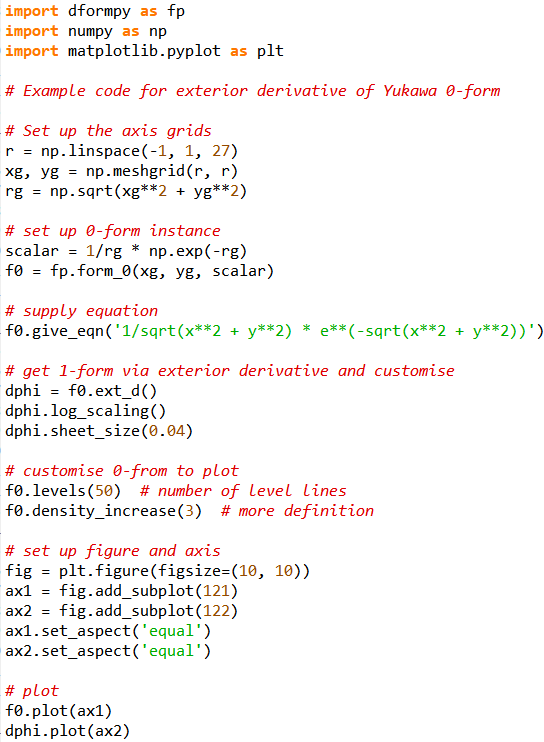}}
        \caption{Example code 2}
        \label{Yukawa_code}
    \end{subfigure}%
    \begin{subfigure}[t]{1\textwidth}
        \centering
        \includegraphics[height=4.35 in]{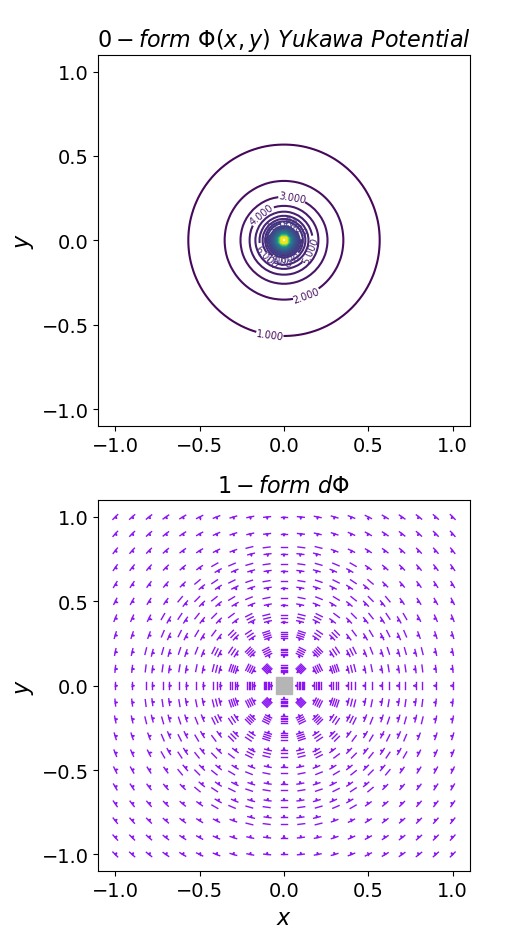}
        \caption{$0$-form and exterior derivative}
        \label{Yukawa}
    \end{subfigure}
    \caption{\fontsize{8}{10}\selectfont 
    Exterior derivative of a $0$-form. The Yukawa potential, $V(r) = e^{-r}/r$ is shown at the top of figure \ref{Yukawa}, using 50 level lines. Lighter colour indicates larger value. Analytic exterior derivative is plotted below it, using logarithmic scaling, on a $27 \times 27$ grid. It demonstrates how the exterior derivative of a $0$-form is analogous to the gradient field. Grey square in the centre represents an occurrence of \inlinecode{Python}{numpy.NaN}, the result of division by zero. 
    }
\end{figure*}
\subsection{Interior derivative}
As discussed in \ref{intd}, the interior derivative method can be used on $2$-forms to create $1$-forms, and on $1$-forms to create $0$-forms. To compute analytically, use the method \inlinecode{Python}{.interior_d( )}, ensuring that expressions for components have been provided. To compute numerically, use \inlinecode{Python}{.num_interior_d( )}.
The vector field can be specified by our vector field instance, a tuple of components grids, or a tuple or equations for said components as strings, depending on user needs (numerical or analytical calculations). If this parameter is omitted, the interior derivative will be taken with respect to $\vec{v} = \hat{x} + \hat{y}$.\\

As an example, we will consider a particle of charge $q$ moving with velocity ${\vec v}$ under the influence of a magnetic field created by a long wire carrying current $I$, and show mathematically and using DFormPy, that the Lorentz force is a $1$-form given by 
\begin{equation}\label{L1}{\pmb F}~=~-q \iota_{\vec v}{\pmb B}.\end{equation}
First, note that the magnetic field is circular in the plane orthogonal to the wire. If we take the wire along the $z-$axis, then the magnetic field is a $2$-form given by ${\pmb B}~=~\frac{\mu_0 I}{2\pi} \frac{dz\wedge d\rho}{\rho}$, where $\mu_0$ is the vacuum permeability. The velocity is given by ${\vec v}=v_\rho\partial_\rho+v_\phi\partial_\phi+v_z\partial_z$, then 
\begin{equation}
{\pmb F} =-q\frac{\mu_0 I}{2\pi} \iota_{(v_\rho\partial_\rho+v_\phi\partial_\phi+v_z\partial_z)}\frac{dz\wedge d\rho}{\rho} =-q \frac{\mu_0 I}{2\pi}\frac{(-v_\rho dz + v_zd\rho)}{\rho}. \end{equation}
We can now switch to the vector version and compare with known result. Since the metric of a cylinder is diagonal with $g_{\rho\rho}=g^{\rho\rho}=1$ and $g_{zz}=g^{zz}=1$, then 
\begin{equation}\label{L2}
{\pmb F} \to {\vec F}~=~q\frac{\mu_0 I}{2\pi}\frac{(v_\rho \hat z - v_z \hat\rho)}{\rho} = q{\vec v} \times {\vec B},\end{equation} where ${\vec v}=v_\rho\hat\rho+v_\phi\hat\phi+v_z\hat z$ and ${\vec B }= \frac{\mu_0 I}{2\pi\rho}\hat\phi$.\\~\\
Now, using DFormPy as shown below, we can verify that \ref{L1} and \ref{L2} match.
\begin{figure}[H]
  \begin{minipage}[c]{0.6\textwidth}
    \includegraphics[width=\textwidth]{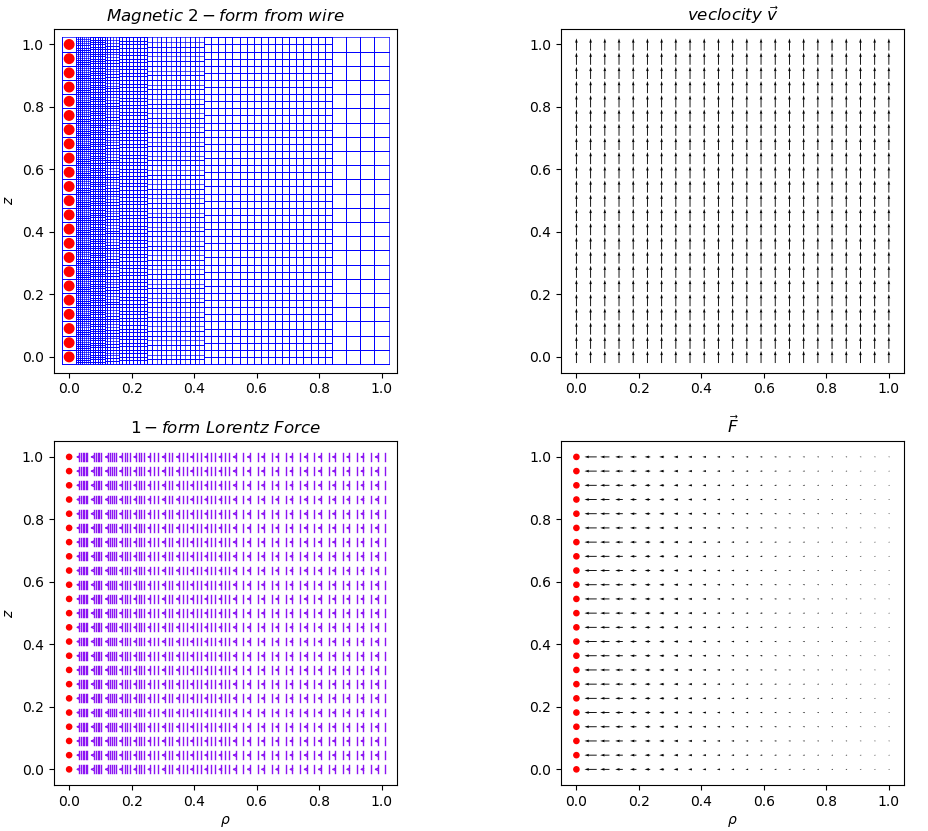}
  \end{minipage}\hfill
  \begin{minipage}[c]{0.4\textwidth}
   \caption{\fontsize{8}{10}\selectfont Lorentz force using the interior derivative and using curl. Figure \ref{fig:Lorentz_Force}a (top left) is the magnetic $2$-form from a wire when viewed side on (wire along vertical axis at $\rho=0$). Figure \ref{fig:Lorentz_Force}b (top right) is the velocity vector field $\vec{v} = \hat{y}$. The interior derivative of the $2$-form w.r.t $\vec{v}$ gives the resulting Lorentz force $1$-form (fig.\ref{fig:Lorentz_Force}c, bottom left). For comparison, the force resulting form the curl $\vec{v} \times \vec{B}$ as a vector is shown on fig.\ref{fig:Lorentz_Force}d. Notice how the results (fig.\ref{fig:Lorentz_Force}c and fig.\ref{fig:Lorentz_Force}d) are the same here, as we working with the flat metric. But the Lorentz force is not to be understood as a vector, since it is covariant in nature and therefore a $1$-form. Logarithmic scaling has been used in all plots to increase the seen variation.}
    \label{fig:Lorentz_Force}
  \end{minipage}
\end{figure}

\subsection{Hodge}
As shown in \ref{hdgfrm}, on $2D$ geometries the Hodge method can be used to change $1$-forms or create $2$-forms from $0$-forms (and vice versa). Use \inlinecode{Python}{.hodge( )} to compute analytically or \inlinecode{Python}{.num_hodge( )} to compute numerically. Methods have one optional parameter, \inlinecode{Python}{keep_object} (applicable where the resulting form is of the same degree as the parent.) If true, the parent $p$-form components will be modified without returning a new object. Otherwise, a new, hodge dual object, will be returned. This parameter is false by default\\
As an example, for the Hodge star operation using DFormPy, we will employ the $2D$ static black hole solution
\begin{equation}
\label{BH}
ds^2=-tanh^2(x)cosh^{4/3}(x)dt^2+dx^2,
\end{equation}
derived in \cite{lemos1995two}. This black hole is a solution of the vacuum planar gravity theory with scalar field and cosmological constant. The solution describes the geometry outside the horizon at $x=0$. The curvature singularity occurs at $\cosh^{2/3}(x)=0$. This means that a coordinate transformation 
\begin{equation}
\label{BHt}
x \to r=\cosh^{2/3}(x),
\end{equation}
gives solution with a black hole like behaviour. If we, however, plot the $1$-forms frame fields of the metric \ref{BH} given by
\begin{equation}
u=dx \quad \text{and} \quad v=tanh(x)cosh^{2/3}(x)dt,
\end{equation}
we expect to get fields with potential that grows as we move away from $x=0$. Using DFormPy, this effect is shown in the figure below, in addition to the effect of the Hodge star operation which inverts the roles of time and space.
\begin{figure*}[ht]
	\centering
	\begin{subfigure}[t]{0.3\textwidth}
		\centering
		\fbox{\includegraphics[height=5 in]{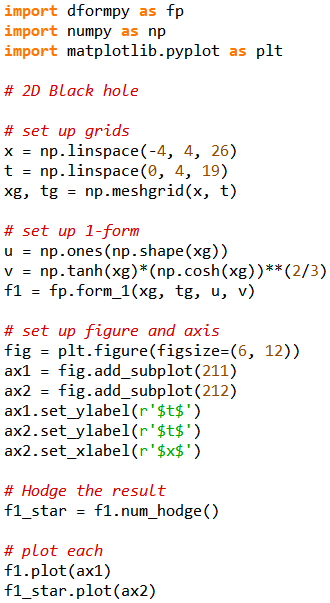}}
		\caption{Example code 3}
		\label{2D_BH_code}
	\end{subfigure}%
	\begin{subfigure}[t]{1\textwidth}
		\centering
		\includegraphics[height=5 in]{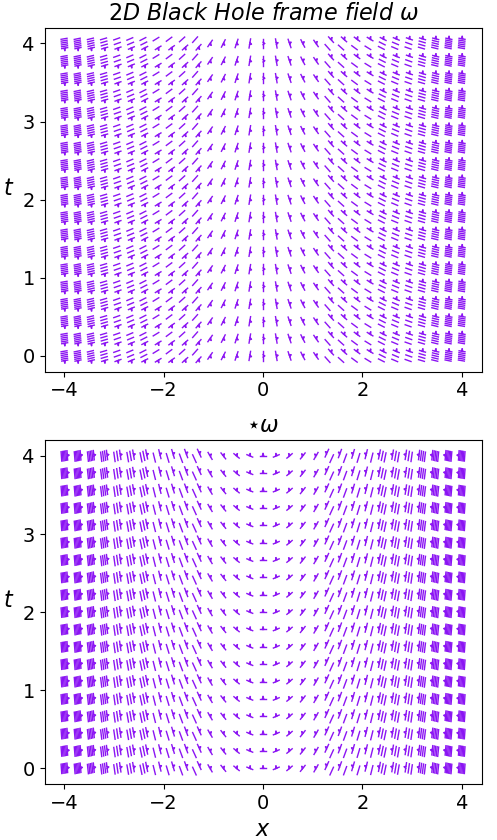}
		\caption{2D Black Hole $1$-form frame fields}
		\label{2D_BH}
	\end{subfigure}
	\caption{\fontsize{8}{8}\selectfont 
		Hodge star of a 2D static black hole frame fields. Top of figure \ref{2D_BH} shows frame fields of 2D static black hole. Bottom, the Hodge star of its frame fields. Note that the use of 2D spacetime is possible in DFormPy if one replaces $(x,y)$ by $(x,t)$, where the extra minus in $dt\wedge*dt=-dV$ is accounted for by the fact that $dV=dt\wedge dx$ in $2D$ spacetime.} 
\end{figure*}
\vspace{-0.5cm}
\begin{wrapfigure}{r}{5cm}
         \captionsetup{width=1.2\linewidth}
	\includegraphics[width=6cm]{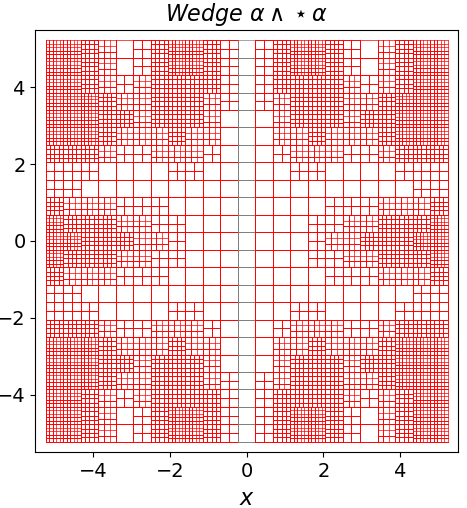}
	\caption{\fontsize{8}{8}\selectfont Using DFormPy's \inlinecode{Python}{.wedge( )} to plot the magnitude square, $\alpha\wedge*\alpha$,  of $\alpha=y\sin(x)dx-x\cos(y)dy$.}
	\label{Wedge:1}
\end{wrapfigure} 
\subsection{Wedge}
The wedge method is used to combine a $p$-form with a $q$-form, creating a $(p + q)$-form. Use \inlinecode{Python}{.wedge( )} to compute analytically or \inlinecode{Python}{.num_wedge( )} to compute numerically. Wedge is found between the parent form and the input form. Inputs can be a tuple of component equations or a DFormPy object with equations provided. For numerical computation, an object, or a tuple of component grids can be supplied. This can be performed between all of our instances. However, created forms of a degree ($p +q$) higher than the the dimension ($2$) are equal to zero. Since the wedge product between two forms is graded commutative \ref{wedgef}, order is important. We choose $a \wedge b$ to be input as \inlinecode{Python}{a.wedge(b)}.
\clearpage
\subsection{Covariant and contravariant}
\label{cotocontra}
The method \inlinecode{Python}{.covariant( )} is used on a vector field object to convert it to a $1$-form. Conversely, \inlinecode{Python}{.contravariant( )} creates a vector field from a $1$-form. Lowering and raising indices involves the metric via $V_i~=~\sum_j  g_{ij}V^j$, and the metric inverse via $V^i= \sum_j g^{ij}V_j$. The user can provide expressions for the metric components or arrays of the evaluated metric components. The default for this parameter is the flat space-metric, where $g_{ij} = diag(1,1)$.
Here, we present an example use of the \inlinecode{Python}{.covariant( )} method.
\begin{figure}[H]
		\centering
	\includegraphics[scale=0.5]{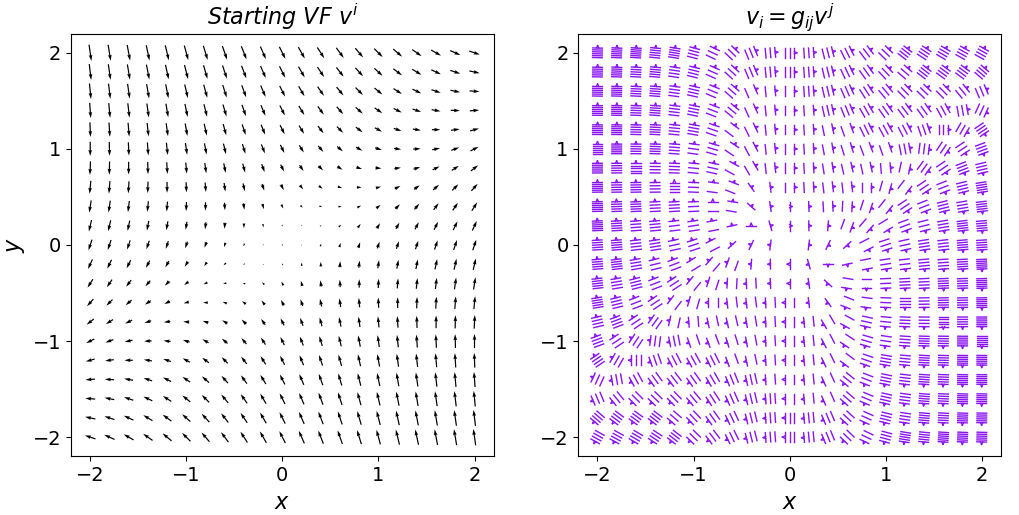}
	\caption{\fontsize{8}{10}\selectfont Using DFormPy's \inlinecode{Python}{.covariant( )} to change a linear contravariant field $\vec{v}=(x+2y)\hat{i}+(3x-4y)\hat{j}$ into a covariant 1-form on a spacetime with the metric of $2D$ black hole given in \ref{BH}.
	}
	\label{fig:form2_and_zooms}
\end{figure}

\section{Conclusion and outlook} \label{Conclusion}
In this paper, we present the main features of DFormPy, the first Python library to provide an interactive visualisation of differential forms and their exterior algebra. In this version of the library we work in ${\mathbb R}^2$, where $0$-forms, $1$-forms and $2$-forms can be used to describe physics. The library is capable of giving representation of any field in ${\mathbb R}^2$ described with differential forms, including fields with singularities, and show any operation on the fields such as exterior derivative, interior derivative and the Hodge star.\\
We believe that this library provides the user with a useful tool to understand differential forms. Many physics books and papers, see for instance \cite{thorne2000gravitation, warnick2014differential, warnick1996differential}, attempt at sketching differential forms to make it easier for the reader to understand them. Sketching differential forms gets very complicated when going beyond linear fields. This library provides this missing tool that renders graphing accurately any differential form on a sheet of paper simple and possible via few Python scripts, with loads of customisation options. The library commands are very intuitive and can be used by Python users of any level. Moreover, at its GitHub,~\url{https://github.com/MostaphaG/Summer_project-df}~, the library is accompanied with a sheet of examples at the TESTS directory, and GUI for further assistance and experience with the library's  commands and features.\\
The uses of the current edition of the library extend from helping with learning and teaching undergraduate modules such as Electromagnetism and Vector Calculus, where a lot of concepts are learnt better with visual examples, to solving spacetime metrics such as black holes metrics. The latter is possible for static and stationary $4D$ black holes where the black holes admit more than two Killing vectors. In these cases, one can construct $1$-form frame fields from the metrics where the weights of the fields are functions of two variables (see for instance last few sections of chapter 3 of \cite{carroll1997lecture} or \cite{krasnov2020formulations}), and hence all the exterior algebra can be done using DFormPy.\\ 
Using DFormPy, users can plot the analytical or the numerical results of the exterior algebra. There is, however, one subtlety regarding fields with closed $1$-forms (conservative vector fields). Since the numerical results for the exterior derivative will shift slightly from the analytical results, it implies that the numerical result of the exterior derivative of a closed $1$-form will not return zero, and hence when plotting its exterior derivative the result will show a non vanishing field in few regions. This is, actually, an issue that we noticed in other numerical vector graphing softwares, and in the attempts to implement the curl for vector fields using NumPy. Our resolution for this issue is making it possible, via two extra commands from DFormPy, to plot as well the analytical result of the exterior derivative to confirm the numerical one if needed.\\
We've already started working on few future additions to the library. A major one is making DFormPy capable of working in ${\mathbb R}^3$, where one can also construct differential $3$-forms. We've done a lot of work on this feature, which will be added soon, leading later to hopefully incorporate time as fourth dimension, and making working in $4D$ spacetime possible via DFormPy. Other small improvements, which we plan to include soon, is extending the ability of the library to handle the metric involvement in the hodge star operation, exterior algebra of forms with complex weights, and weights that are functions of  more than two variables. These additions will make it possible for the user to do the exterior algebra of  Dirac spinors in $4D$ spacetime, where spinors can be represented as complexified differential forms on ${\mathbb R}^2$ \cite{lawson2016spin}.\\
Like any other Python package, DFormPy is available at PyPI,~\url{https://pypi.org/project/dformpy/}~. The package can be installed using \inlinecode{bash}{pip install dformpy}. Once installed, all the modules that DFormPy depends on will also be installed. For most of DFormPy coding, as shown in \ref{form_1_zoom_code}, \ref{Yukawa_code}  and \ref{2D_BH_code}, the user needs to import numpy and matplotlib.pyplot along with formpy. Moreover, as mentioned earlier, DFormPy project, at its GitHub, has a user friendly GUI. The GUI is meant to give an interactive demonstration for all the functionalities of DFormPy without the need for any coding. Users, via the GUI, only need to enter the differential form or the vector field components and click to visualise their geometric structures, their calculus and exterior algebra or their time evolutions. In the GUI, we have also implemented some extra tools, that are still under testing, such as visualising differential forms on $\mathbb{R}^3$. These extra tools will be added to DFormPy in the near future, once tests and analyses are completed.\\~\\ 
{\bf{Acknowledgements}}: This paper is based on the summer 2021 internship project supervised by the first author. MJ and SKJ are grateful to the School of Physics and Astronomy at the University of Nottingham for supporting their work during the internship.\\~\\


\appendix

\section{Customisations} \label{appendix_custom}
\subsection{$1$-Form customisations}
\label{A1}
\begin{itemize}
	\item \inlinecode{Python}{.colour( )}: changes colour of stacks, parameter: matplotlib colour understood string, or Hex colour code.
	\item \inlinecode{Python}{.arrow_heads( )}: enable/disable stack arrowheads , no parameter
	\item \inlinecode{Python}{.head_width( )}: length of the arrowhead base as a fraction of stack length, parameter: positive float $\leq 1$
	\item \inlinecode{Python}{.head_height( )}: length of the arrowhead (base to tip) as a fraction of stack length, parameter: positive float $\leq 1$
	\item \inlinecode{Python}{.log_scaling( )}: enable/disable logarithmic scaling of the number of stacks, no parameter
	\item \inlinecode{Python}{.max_sheets( )}: maximum number of sheets per stack, parameter: positive integer
	\item \inlinecode{Python}{.sheet_size( )}: length of sheet as a fraction of the plot size, parameter: positive float $\leq 1$
	\item \inlinecode{Python}{.surround_space( )}: Changes width of empty boarder around plotted region as the fraction denominator of total plot size, parameter: integer or float $\geq 1$
	\item \inlinecode{Python}{.set_density( )}: recalculate the form components on $n \times n$ grid, over the same range as grids provided by the user, parameter: positive integer $n$. \textbf{Important}: only works if instance contains equations.
\end{itemize}

\subsection{Vector field customisations}
\label{A2}
\begin{itemize}
	\item \inlinecode{Python}{.colour( )}, \inlinecode{Python}{.log_scaling( )}, \inlinecode{Python}{.surround_space( )}, \inlinecode{Python}{.set_density( )}:
	\item \inlinecode{Python}{.orient( )}: Changes orientation of arrows, parameter: string same as matplotlib's \inlinecode{Python}{.quiver()} orientation input.
	\item \inlinecode{Python}{.autoscale( )}: Auto-scales the arrows relative to the maximum magnitude, no parameter.
\end{itemize}

\subsection{$0$-Form customisations}
\label{A3}
\begin{itemize}
	\item \inlinecode{Python}{.surround_space( )}, \inlinecode{Python}{.set_density( )}:
	\item \inlinecode{Python}{.density_increase( )}: If strings are supplied, changes the density of points same as \inlinecode{Python}{.set_density( )}, but only for plotting purposes, parameter: factor of density increase, positive integer. \textbf{Important}: only works if instance contains equations.
	\item \inlinecode{Python}{.levels( )}: defines number of level lines to draw, parameter: positive integer or (ascending) list passed to contour via levels parameter. If integer, matplotlib automatically sets level values, if list, values set to ones in list.
	\item \inlinecode{Python}{.labels( )}: Changes boolean that determines if labels are drawn on level lines, no parameter
	\item \inlinecode{Python}{.fonts_size( )}: Changes font of labels, if labels are set to True, parameter: positive integer. 
\end{itemize}

\subsection{$2$-Form customisations}
\label{A4}
\begin{itemize}
	
	\item \inlinecode{Python}{.log_scaling( )}, \inlinecode{Python}{.max_sheets( )}, \inlinecode{Python}{.sheet_size( )}, \inlinecode{Python}{.surround_space( )},
	\item \inlinecode{Python}{.set_density2( )} recalculates the form components on $n \times m$ grid, over the same range as grids provided by the user, parameters: positive integer $n$, positive integer $m$. \textbf{Important}: only works if instance contains equations.
	\item \inlinecode{Python}{.colours( )}: changes the set of colours used for $2$-form orientations, parameter: list of strings, format [counter-clockwise, clockwise, zero]
\end{itemize}

\subsection{Singularities}
\label{A5}
The use of relative scaling in many DFormPy methods, makes it susceptible to breakdown from singularities that occur on evaluated grid points. To stop this from rendering our methods helpless, we have implemented a simple search algorithm, capable of identifying divergent and ill-defined points. To do this, we check against any point being evaluated as not-a-number (\inlinecode{Python}{np.NaN}), an infinity (\inlinecode{Python}{np.inf}) and/or a value that NumPy may evaluate as approximately infinite by making it larger than $1\times 10^{15}$. In each case, we set the corresponding value for the scaling function to zero, and in its position we plot a marker. We define our makers based on the NumPy evaluation. Points containing (\inlinecode{Python}{np.inf}) or values larger than $1\times 10^{15}$ are marked using a red circle, while those containing \inlinecode{Python}{np.NaN} are shown with a grey square.

\subsection{Log scaling}
\label{A6}

In the event that the field varies over multiple orders of magnitude, the user can use the \inlinecode{Python}{.log_scaling()} method to display variation in the field more clearly. This will recalculate the field components by normalising $u$, $v$ and $w$ grid values, then multiplying by the logarithm of the magnitude at each point. This version of DFormPy uses logarithms with base 10. 

\section{A Note on $2$-forms visualisation}
Using DFormPy, $2$-forms can be provided directly by the user, or generated by taking the exterior derivative of $1$-forms, taking the wedge product of two $1$-forms or taking the Hodge star of a $0$-forms.\\
 A graphical representation of a $2$-form must be unique regardless of how it's generated. During the early days of this project, we represented a given $2$-form by superposing two $1$-forms stackings. This works well if the $2$-form is a result of wedge product of two $1$-forms. However, it doesn't provide a unique graphical representation for $2$-forms generated by any other method. For instance, given a $2$-form, which can be a user input, one can find an infinite number of two $1$-forms where their wedge product gives that particular $2$-form. Hence, if we rely on the superposition of the stackings of the underlying couples of $1$-forms, we will end up with infinite number of graphical representations of the same $2$-form.\\
The main issue in that approach is that we were trying to do the graphical representation before the algebra. That left us with a problem of how to translate the algebra (wedge, Hodge star or exterior derivative) graphically. The resolution was to do the graphical representation after the algebra, which implies using the weights of the $2$-forms to represent them. By that we mean, evaluating the $2$-form weight at each coordinate in the grid, then to use its magnitudes as the squares areas and its signs as the squares colour, red if positive (counter-clockwise) and blue if negative (clockwise). 

\section{Geometric derivative, divergence and curl}
The vector fields' geometric derivative, divergence and curl methods have been adopted from the JAVA code used in the VFA. The JAVA code is available online via \cite{VFA2020}. The name of the main code file for VFA is GraphCanvas.java. The GraphCanvas.java function that deals with these operations is called \inlinecode{JAVA}{public void Mag4(vfa2 app)}. We studied this function, and the math used, and code it in Python for DFormPy.

In short, the main idea is that when looking for the derivative of a vector field locally, around certain point, one can take the field at that point as a background field for the local region of interest. Then, in the local region, subtract the background field from the field at each point. This gives the Lie derivative vector field locally at each point. For the curl and the divergence, one has to extract them from the derivative field information at each point. This can be done by using the infinitesimal generator of the plane rotation group, $SO(2)$, to write the derivative on a vector field as operator. This allows us to identify the commuting and anti-commuting part of the derivative operator, which can be used to act on each point in the local region. This action generates two matrices, one has the values of the commuting part of the derivative vector field at each point, and the other has values for the anti-commuting part of the derivative. Then, by projection, the tangential coefficients of the commuting operator values give the curl field at each point, and the normal coefficients give the divergence field at each point.


%
%
%
%
%

\bibliographystyle{utphys}
\bibliography{paperdf}

\end{document}